\documentclass[debug]{rmaa}

\usepackage{paralist}

\usepackage{psfrag,color}

\usepackage[latin1]{inputenc}

\title{A mid infrared study of low-luminosity AGNs with WISE} 

\author{
R. Coziol,\altaffilmark{1}
J. P. Torres-Papaqui,\altaffilmark{1} 
I. Plauchu-Frayn,\altaffilmark{2}
H. Andernach,\altaffilmark{1}
D. M. Neri-Larios,\altaffilmark{3}
R. A. Ortega-Minakata,\altaffilmark{1}
and J. M. Islas-Islas \altaffilmark{1}}

\altaffiltext{1}{Departamento de Astronom\'{\i}a, Universidad de Guanajuato, Apartado Postal 144, 36000, Guanajuato, Gto, Mexico.}

\altaffiltext{2}{Instituto de Astronom\'{\i}a, Universidad Nacional Aut\'onoma de M\'exico, Campus Ensenada, Apdo. Postal 877, Ensenada, B.C.N. Mexico.}

\altaffiltext{3}{School of Physics, The University of Melbourne, Parkville, Vic. 3010, Australia.}

\shortauthor{Coziol et al.}
\shorttitle{Low-luminosity AGNs in MIR}

\fulladdresses{
\item R. Coziol, J. P. Torres-Papaqui, H. Andernach, R. A. Ortega-Minakata and J. M. Islas-Islas: Departamento de Astronom\'{\i}a, Universidad de Guanajuato, Apartado Postal 144, 36000, Guanajuato, Gto, Mexico (rcoziol, heinz, papaqui \& rene@astro.ugto.mx).
\item I. Plauchu-Frayn: Instituto de Astronom\'{\i}a, Universidad Nacional Aut\'onoma de M\'exico, Campus Ensenada, Apdo. Postal 877, Ensenada, B.C. Mexico (ilse@astrosen.unam.mx).
\item D. M. Neri-Larios: School of Physics, The University of Melbourne, Parkville, Vic. 3010, Australia (danielnl@student.unimelb.edu.au).
}

\listofauthors{J. P. Torres-Papaqui,  R. Coziol, I. Plauchu-Frayn, H. Andernach, \& R. A. Ortega-Minakata}
\indexauthor{Coziol, R.}
\indexauthor{Torres-Papaqui, J.P.}
\indexauthor{Plauchu-Frayn, I.}
\indexauthor{Andernach, H.}
\indexauthor{Neri-Larios, D. M.}
\indexauthor{Ortega-Minakata, R.A.}
\indexauthor{Islas-Islas, J. M.}

\abstract
{Using data from the Wide-field Infrared Survey Explorer (WISE) we show that the mid infrared (MIR) colors of
low-luminosity AGNs (LLAGNs) are significanlty different from those of post-asymptotic giant branch stars (PAGBs). This is due to a difference in spectral energy distribution (SEDs), the LLAGNs showing a flat component due to an AGN. Consistent with this interpretation we show that in a MIR color-color diagram the LINERs and the Seyfert~2s follow a power law with specific colors that allow to distinguish them from each other,  and from star forming galaxies, according to their present level of star formation. Based on this result we present a new diagnostic diagram in the MIR that confirms the classification obtained in the optical using standard diagnostic diagrams, clearly identifying LINERs and LLAGNs as genuine AGNs.}

\resumen
{Utilizando datos en el infrarrojo mediano (MIR) del Wide-field Infrared Survey Explorer (WISE), mostramos que las 
galaxias AGNs de baja luminosidad (LLAGNs) presentan colores en el infrarrojo mediano (MIR) significamente diferentes de
estrellas post-rama asint\'otica gigante (PAGBs). Esto es debido a una diferencia en la distribuci\'on de energ\'{\i}a espectral (SED), los LLAGNs mostrando una componente plana debida a un AGN. Consistente con esta interpretaci\'on, mostramos que en un diagrama color-color los LINERs y las Seyfert~2s siguen una ley de potencia con colores espec\'{\i}ficas que permiten distinguir unas de las otras, y de las galaxias con formaci\'on estelar, en base a sus diferentes niveles de formaci\'on reciente de estrellas. Basado en estos resultados, presentamos un nuevo diagrama de diagn\'ostico en el MIR que confirma la clasificaci\'on obtenida en el \'optico a partir de diagramas de diagn\'ostico
estandard, identificando claramente los LINERs y LLAGNs como verdareros AGNs.}

\addkeyword{infrared: galaxies}
\addkeyword{galaxies: active}

\begin{document}
\maketitle

\section{Introduction}
\label{sec:intro}

Studies of clusters and compact groups \citep{Phi86,Coz98,Mil03,Mar08,Mar10} have revealed
that many narrow emission-line galaxies (NELGs) cannot be classified using standard
diagnostic diagrams \citep{BPT81,VO87}, because the most important emission lines,
H$\beta$ and [OIII]$\lambda5007$, are either too weak or not observed. This is rather
common, affecting about 20\% of the early-type galaxies in compact groups, but reaching
60\% in clusters \citep{Phi86,Coz98,Mil03,Mar08}. 

After subtracting a stellar population
template from their spectra, \citet{Coz98} have demonstrated that these unclassified NELGs
have spectral characteristics typical of low luminosity AGNs: the emission lines have
small equivalent widths (EWs), consistent with low emission luminosities, and the ratios
[NII]$\lambda6584/$H$\alpha$ are high, which is a
defining trait of AGNs \citep{BPT81,VO87,Ost89}. Subsequently, \citet{Mar08} confirmed
that even after correcting for dust extinction the median value of the H$\alpha$
luminosity in these galaxies is only $7.1\times 10^{39}$ erg s$^{-1}$, which qualify them
as low-luminosity AGNs \citep[LLAGNs;][]{HFS97,ZDW07,Mar08,Mar10}.

The most straightforward interpretation for LLAGNs is that they are some sort of evolved
AGNs or ``dying quasars'', where matter at the center of the galaxies is falling on the
exhausted accretion disk of a super-massive black hole (BH), reviving its activity
\citep{Coz98,Ric98,Mil03,Gav08}. However, an alternative interpretation, put forward by
\citet{Sta08} and \citet{Cid10,Cid11}, suggests these galaxies are ``retired galaxies'',
falsely identified as AGNs, in which the gas is ionized by post-asymptotic giant branch
stars \citep[PAGBs;][hereafter the PAGB hypothesis]{Bin94}. During the last few years,
the PAGB hypothesis has won in popularity and was even proposed
\citep[e.g.][]{EHF10,Singh13} to explain the LINERs, the prototype LLAGN that forms about
30\% of all the early-type spiral galaxies observed in the field \citep[][]{Hec80,Kau09}.

However, the PAGB hypothesis is somewhat vague and ambiguous. The whole evolution from the
AGB to the white dwarf cannot last more than 10000 years \citep{VolkKwok89}. This is a
transient phenomenon that does not fit well with the ubiquity of the LLAGNs. It is not
clear either to what phase of the post AGB evolution the PAGB hypothesis is alluding to.
By definition, an AGB is a red giant star that does not produce ionizing photons. When this
star collapses, transforming into a white dwarf, the central star, for a brief period,
becomes extremely hot, its effective temperature reaching up to 10$^5$~K
\citep{McCookSion99,Eisenstein06}. Sometime during this transformation, the star emits
ultraviolet photons that are responsible for ionizing the gas of the envelope recently
ejected from the AGB. This phase corresponds to that of the planetary nebula (PN).
After that, the effective surface temperature of the central star falls down rapidly to
about $10^4$~K, which is not hot enough to ionize the gas. The expanding gas cloud thus
becomes invisible, ending the PN phase \citep{Kwok00}. The star is now a white dwarf.

Strictly speaking, therefore, a PAGB is pre-planetary nebula \citep[pre-PN; see definition
in][]{Davis05,Sua06,Kitsikis07}, which describes the stage when the temperature of the
evolving AGB is still rising, and that lasts only a few thousand years \citep{VolkKwok89}.
Sometime during this brief phase the UV photons from the hot white dwarf would be able to
escape from the expanding gas and dust envelope of the AGB and ionize the interstellar gas
over a kpc region. The exact mechanism by which this is possible remains, however, largely
unexplained \citep{Bin94}. Ignoring this problem, \citet{Taniguchi00} claimed that it is
not the pre-PN stars that ionize the gas in a galaxy, but the central hot white
dwarfs of the PNs themselves. However, this would imply that the gas nebulae are generally
optically thin, which means that all the PNs are density-bound. This is an
oversimplification of the PN theory \citep[see the basic explanations in][]{OF06}.
Moreover, this complicates the PAGB hypothesis, because one needs then to establish what
fraction of the ionizing photons are escaping the nebulae. Consequently, the fluxes of UV
photons required to ionize the gas might not be sufficient to explain a typical LLAGN.
This is unless we have a very high number of PNs at all times, which is highly improbable
considering the short lifetime of this phenomenon.

The above considerations did not deter \citet{Sta08},  who claimed that
both PAGBs (it is not clear if the authors refer to pre-PNs, PNs or both) and white dwarfs
are responsible in producing the photons that can ionize the gas in LLAGNs over kpc
regions. The ad hoc addition of white dwarfs to the PAGB hypothesis is interesting. White
dwarfs can be present in great numbers in very old galaxies, which is the case of most
LLAGNs \citep{TP13}, and they can live for a very long time. However, strictly speaking,
we would call this model the ``hot white dwarfs hypothesis'', because only white dwarfs with
an effective temperatures above 18000~K can produce ionizing photons \citep{Bianchi11}.
Again, this restriction might significantly reduce the available flux of
ionizing photons in a LLAGN, considering that the majority of white dwarfs are cold, having a surface
temperature well below 18000 K \citep{McCookSion99,Eisenstein06,Bianchi11}.

Taking into account the three possible ``post AGB'' phases as described above, there is one clear and immediate
consequence of the PAGB hypothesis, which is that one would need a very high number of ``hot PAGBs'' to ionize the
amount of gas observed in a typical LLAGN. In \citet[][see their Table~1]{Bin94}, the 15 galaxies for which the PAGB
hypothesis was proposed have a mean H$\alpha$ luminosity of $10^{39.6}$\ erg s$^{-1}$ (converting their cosmology to the
one we used). In \citet{Mar08} an average luminosity of $10^{39.8}$\ erg s$^{-1}$ is reported for the LLAGNs (about 100
galaxies) in compact groups. In comparison, the 43922 LLAGN candidates in our present sample have an H$\alpha$
luminosity that varies between $10^{41}$ and $10^{39}$\ erg s$^{-1}$. Adopting $10^{40}$\ erg s$^{-1}$ for the mean
H$\alpha$ luminosity and using the equation $Q_H \sim 7.3 \times 10^{11}$L$_\alpha$\ photon s$^{-1}$ for the relation
between the flux of ionizing photons, $Q_H$, and the H$\alpha$ luminosity, L$_\alpha$
\citep{Pottasch65,Kennicutt94,Madau98}, we deduce that a flux of $10^{51.9}$ photon s$^{-1}$ is needed to explain the
ionized gas in a typical LLAGN. In Figure~1 of \citet[][]{Bin94} the normalized flux of ionizing photons produced by the
PAGBs after about 13 Gyrs is predicted to be $10^{41}$\ photon s$^{-1}$\ M$^{-1}_\odot$, which gives a mass of stars
equal to $10^{51.9}/10^{41} = 10^{10.9}$\ M$_\odot$. Since the central star of a PAGB is a hot white dwarf with a mass
$\sim 1$\ M$_\odot$, this implies that of the order $10^{11}$ hot PAGBs are needed to ionize the gas in a typical LLAGN.
Note that one obtains exactly the same number of white dwarfs using the model of \citet{Sta08}. Therefore, between
$10^{12}$ and $10^{10}$ hot white dwarfs would be required by the PAGB hypothesis to explain the ionized gas in the
LLAGN candidates forming our sample.

The fact that we need such a high number of white dwarfs is easy to understand. According
to \citet{CBD95} the equivalent of $10^5$ O and B stars would be needed to ionize the gas
in an average LLAGN \citep[note that the same number of O and B stars is
predicted after 10$^7$ yrs by the model of][]{Bin94}. Because the number of ionizing
photons emitted by a star is directly proportional to its surface area, which for a
typical white dwarf is only $10^{-6}$ times that of a normal O star, then of the order of
$10^5/10^{-6} = 10^{11}$ white dwarfs are thus required.

Obviously, like the massive stars in the SFGs, such a high number of hot white dwarfs in the
LLAGNs must leave their trace elsewhere than in the optical spectra. In particular, it
should be possible to test directly the PAGB hypothesis for the LLAGNs in the mid infrared
\citep[MIR; e.g.][]{Pas00}. The MIR emission is due to dust, which in a galaxy is
either heated by stars or an AGN \citep{Wright10,Jarrett11,Mat13,Assef13}. Since these two
sources have different spectral energy distributions (SEDs), stars being similar to black
bodies and AGN emitting energy through a power-law \citep{AlonsoHerrero06}, the light
re-emitted by dust should thus have different SEDs. Consequently, a high number of hot
PAGBs in a LLAGN must produce characteristic colors in the MIR that should be clearly
distinguished from those produced by an AGN.

In their study, \citet{Pas00} observed that some early-type galaxies with weak emission
lines have a ``blue'' SED in the optical and infrared. They consequently suspected that
either an AGN or PAGBs produced these blue SEDs. However, based on their MIR data they were
unable to distinguish between a black body and a power law. Nonetheless, these authors
concluded in favor of the PAGB hypothesis because they did not observe an excess of
infrared emission, which they assumed must be a characteristic of any AGNs in the MIR. However, the
sample in \citet{Pas00} was too small (only 28 galaxies) to establish what are the
``normal'' MIR characteristics of NELGs with different activity types, and these authors
did not compare the MIR emission of their galaxies with the actual MIR emission produced
by PAGB stars. These are deficiencies that can be easily rectified today thanks to data from the
Sloan Digital Sky Survey \citep{York00} and the Wide-field Infrared Survey Explorer
\citep[WISE;][]{Wright10}.

 \section{Selection of the samples}

\subsection{The ``standard'' NELGs and the LLAGN candidates}

\begin{figure}[!t]
\begin{center}
\includegraphics[width=0.9\textwidth]{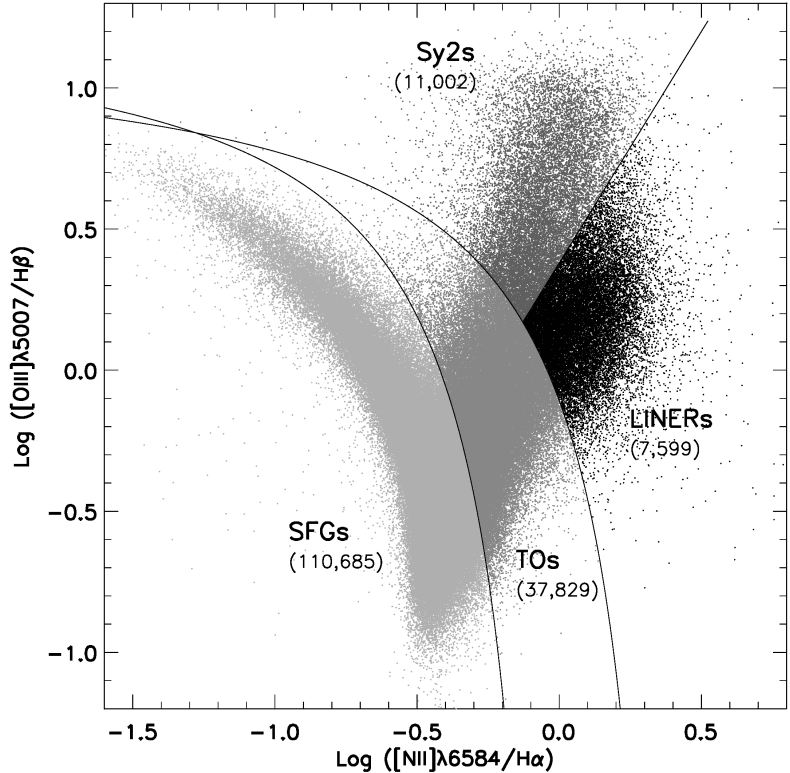}
\caption{The BPT-VO diagnostic diagram for the SDSS standard NELGs. The
separation between SFGs and TOs was suggested by \citet{Kau03} and the
separation between AGNs and TOs by \citet{Kew01}. The separation between the LINERs and Sy2s was established in \citet{TP12a}.}
\label{fig:1}
\end{center}
\end{figure}

To establish what are the normal characteristics of galaxies with different activity types
in the MIR, we have assembled a large sample of ``standard'' NELGs, and searched for
corresponding data in WISE. The ``standard'' NELGs are defined as NELGs that can be
classified using standard diagnostic diagrams. These standard NELGs were thoroughly
studied by \citet{TP12a}. In Fig.~1 we show one standard diagnostic diagram
\citep[hereafter, the BPT-VO diagram;][]{BPT81,VO87}, that we used to separate the
standard NELGs into star forming galaxies (SFGs), transition type objects (TOs), Seyfert 2
(Sy2s) and LINERs. To distinguish between LINERs and Sy2s we followed the method described
by \citet{Kew06}. The distinction between the two activity types is based on differences
in the [OI]$\lambda6300$ and [SII]$\lambda\lambda6717,6730$\ emission-line intensities
\citep{TP12a}.

The NELGs for our study were selected from the main spectral catalog of the Sloan Digital
Sky Survey Data Release~7 (SDSS DR7) \citep{Aba09}, which represents the completion of the
SDSS project, a series of three interlocking imaging and spectroscopic surveys, carried
out over an eight-year period with a dedicated 2.5m telescope located at Apache Point
Observatory in Southern New Mexico. The SDSS DR7 catalog includes the spectra of $9.3
\times 10^5$\ galaxies, $1.2 \times 10^5$ quasars and $4.6 \times 10^5$\
stars\footnote{http://www.sdss.org/dr7/start/aboutdr7.html}.

As a first selection criterion we kept only the galaxies with redshift $z \le~0.25$. Then,
after correcting for the redshifts and subtracting stellar templates produced by STARLIGHT \citep{Cid05}, we applied the
signal-to-noise (S/N) criteria adopted by \citet{Brinchmann04}, \citet{Kew01,Kew06}, \citet{Kau03} and
\citep{Cid10}: we kept only the galaxies that have line ratios in emission with ${\rm
S/N} \ge 3$, and ${\rm S/N} \ge 10$ in the adjacent continuum.

For the MIR data, we cross-correlated the positions of the NELGs in our sample with
objects in the catalog produced by WISE \citep{Wright10}, which is available through the
IRSA (IR Science
Archive)\footnote{http://irsa.ipac.caltech.edu/cgi-bin/Gator/nph-scan?mission=irsa\&submit=Select\&projshort=WISE}.
Using a radius of one arcsecond around the positions of the galaxies, our search produced
110685 SFGs, 37829 TOs, 11002 Sy2s and 7599 LINERs. The results of our search were
confirmed independently using the X-Match pipeline in VizieR \citep{OBM00}. Our selected
galaxies have WISE fluxes with signal to noise $S/N > 2$ (corresponding to quality flags
ph\_qual equal to A, B or C) in all of the four MIR bands of WISE (3.3, 4.6, 12, and 22
$\mu$m).

To test the PAGB hypothesis we have also selected from SDSS DR7 a sample of LLAGN
candidates. In \citet{Coz98} the LLAGNs were characterized by their spectra as galaxies
that show, after a stellar template subtraction,  high ratios of
[NII]$\lambda6584/$H$\alpha$ typical of AGNs, but that cannot be classified using
standard diagnostic diagrams, because either H$\beta$ or [OIII]$\lambda5007$, or both
emission lines are undetected.

From the SDSS DR7 spectroscopic catalog we first selected 476841 NELGs with a ${\rm S/N} \ge 10$ in
the continuum and redshift $z \le 0.25$. After having corrected for the redshifts and subtracted stellar templates
produced by STARLIGHT, we found 10926 galaxies with both H$\beta$ and [OIII] undetected
(hereafter identified as LLAGN$\alpha$), 20784 galaxies with H$\beta$ detected but
[OIII]$\lambda5007$ undetected (identified as LLAGN$\beta$), and 61290 galaxies 
with [OIII] detected but H$\beta$ undetected (identified as LLAGN$\gamma$). Together,
these three samples represent about 20\% of the whole sample of NELGs. Keeping only the
NELGs that have a ${\rm S/N} \ge~3$ in the remaining emission lines reduces the sample to
219375 galaxies (46\% of the original sample). In this sample we distinguish 4198
LLAGN$\alpha$, 12378 LLAGN$\beta$ and 36524 LLAGN$\gamma$, which represents $\sim24$\% of the
NELGs with ${\rm S/N} \ge~3$. After cross-correlating the positions of these galaxies
with those of the objects in the WISE catalog, applying the same search radius and
quality criterion on the fluxes as for the standard NELGs, we found 2840 LLAGN$\alpha$,
9615 LLAGN$\beta$, and 31467 LLAGN$\gamma$ (83\% of the LLAGN candidates with ${\rm S/N}
\ge 3$). Counting only the NELGs with WISE data (176115 standard NELGs and 43922 LLAGN
candidates), the NELGs with undetected emission represent 21\% of the whole sample. Contrary to what was claimed in \citet{Cid10}, the fact
that the fraction of NELGs with undetected lines stays constant suggests that the
reason why some emission lines are not detected in these galaxies is not due to a low S/N (see discussion in Section~2.2 below).

\subsection{Classification of LLAGNs according to the WHAN diagnostic diagram}

\begin{figure}[!t]
\begin{center}
\includegraphics[width=0.7\textwidth]{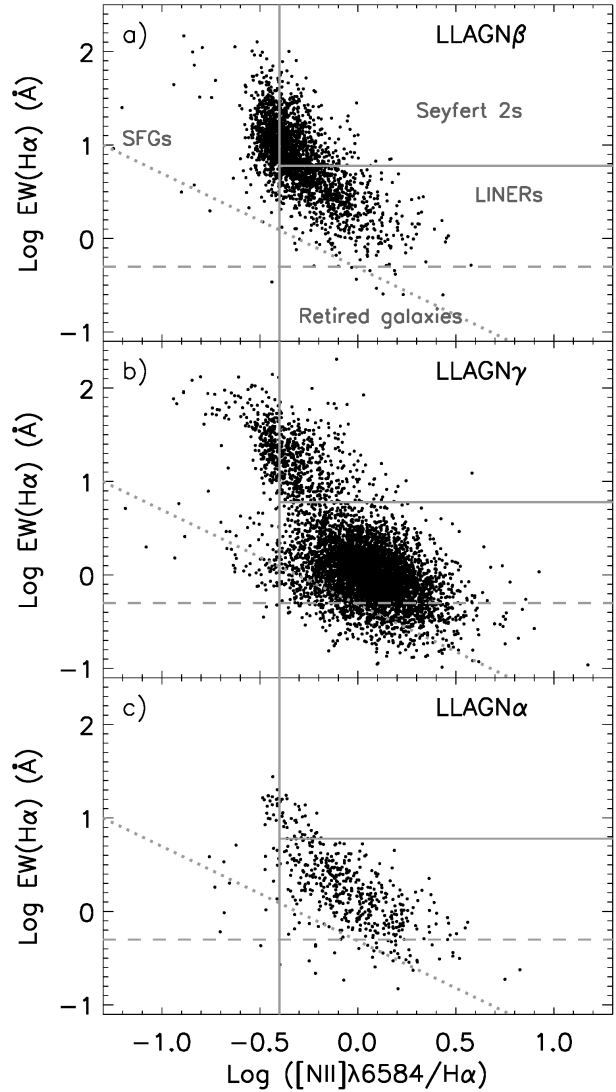}
\caption{WHAN diagnostic diagrams for our three samples of NELGs with some emission lines
undetected: a) 9615 LLAGN$\beta$, b) 31467 LLAGN$\gamma$, and c) 2840 LLAGN$\alpha$ (see descriptions in the text). The separations between the activity types are as defined in \citet{Cid10}.}
\label{fig:2}
\end{center}
\end{figure} 

\begin{figure}[!t]
\begin{center}
\includegraphics[width=0.7\textwidth]{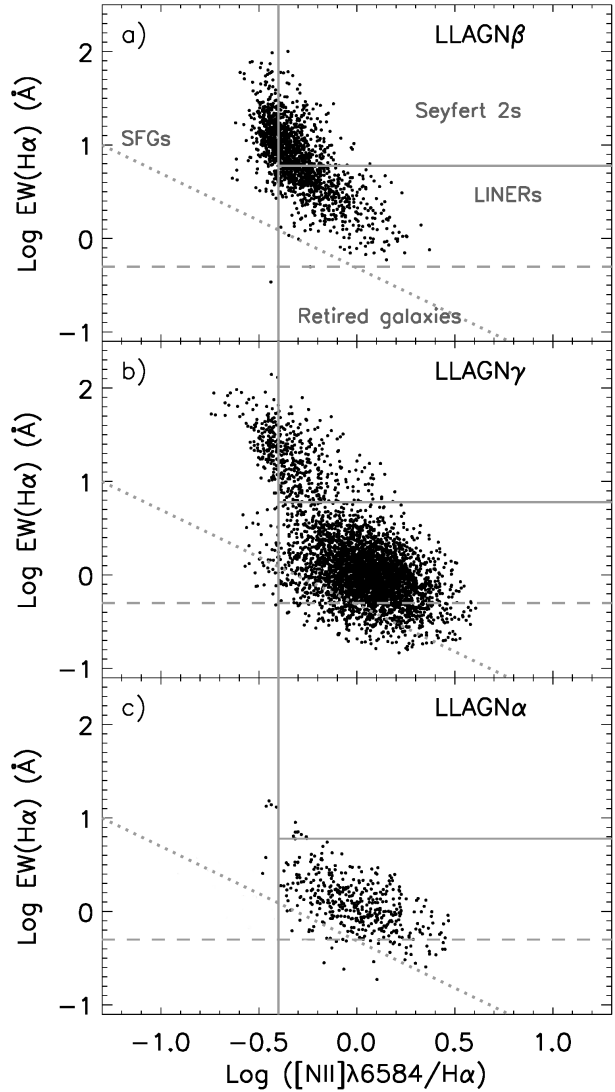}
\caption{Same as Fig.~2 keeping only the NELGs with S/N $\ge 5$.}
\label{fig:2}
\end{center}
\end{figure}  

\begin{figure}[!t]
\begin{center}
\includegraphics[width=0.7\textwidth]{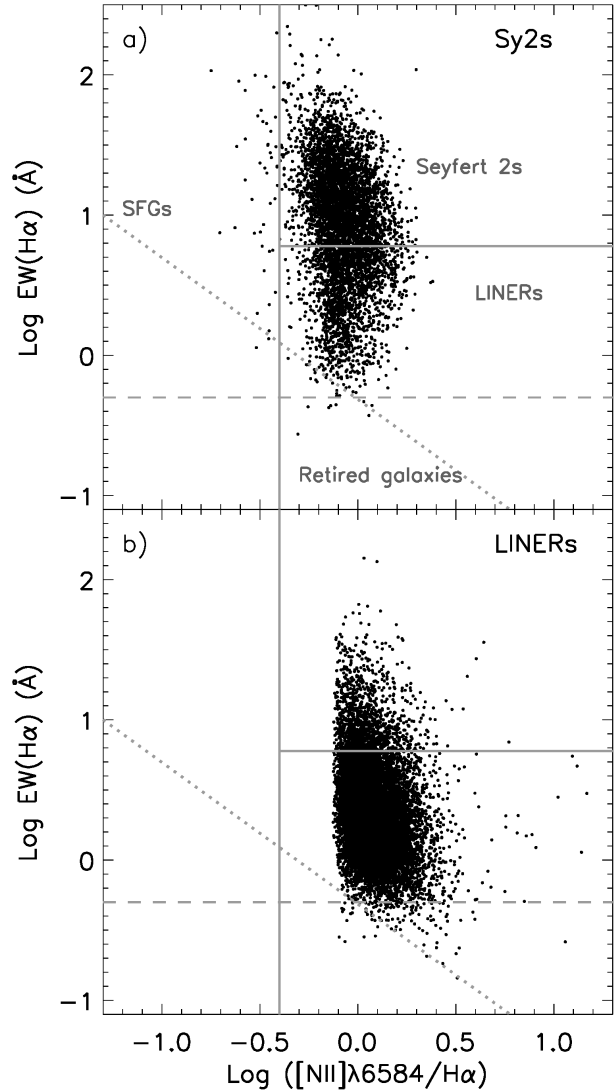}
\caption{WHAN diagnostic diagram for the standard NELGs: a)
11002 Sy2s and b) 7599 LINERs. }
\label{fig:4}
\end{center}
\end{figure}

\begin{figure}[!t]
\begin{center}
\includegraphics[width=0.7\textwidth]{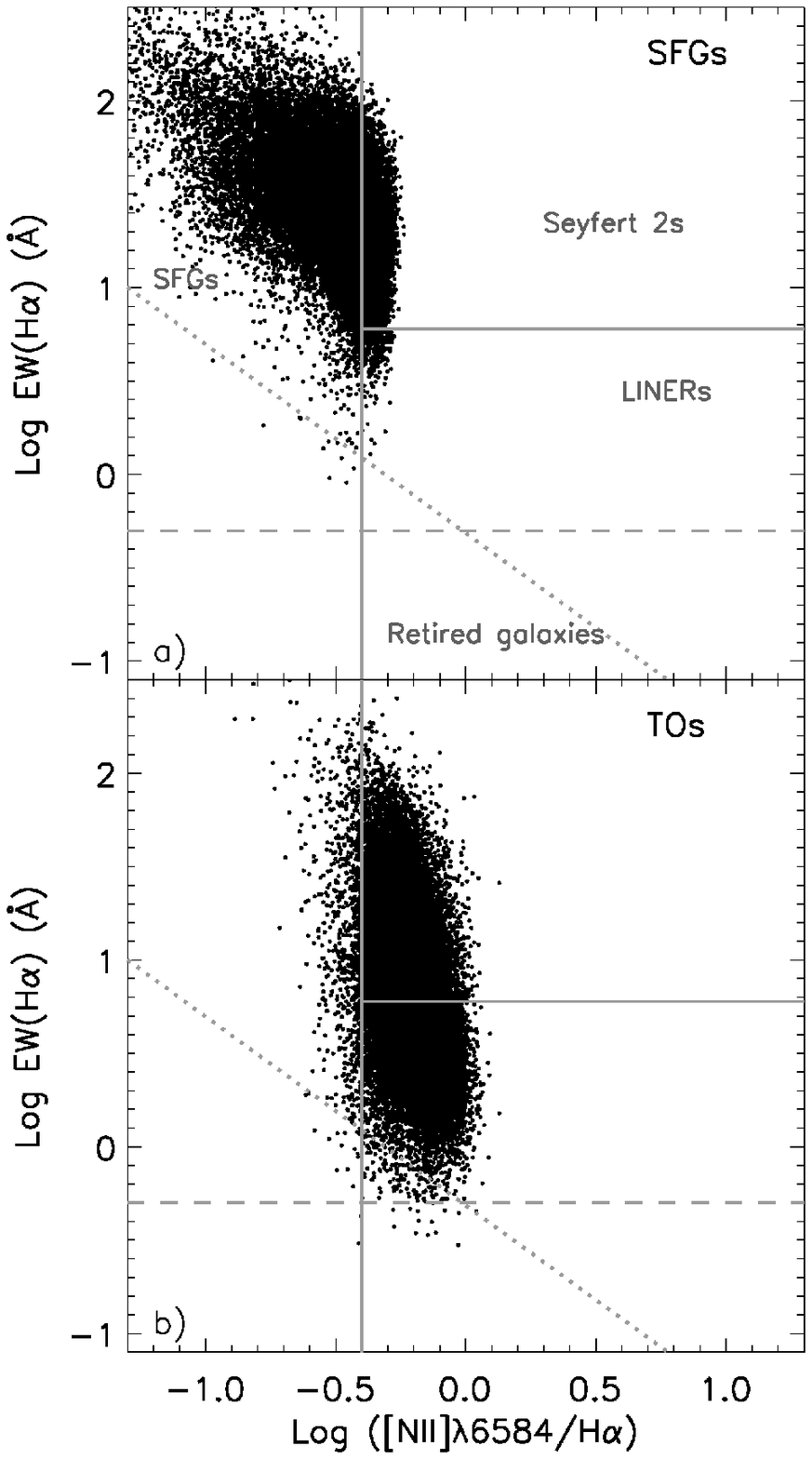}
\caption{WHAN diagnostic diagram for the standard NELGs: a)
110685 SFGs and b) 37829 TOs. }
\label{fig:5}
\end{center}
\end{figure}

According to \citet{Cid10,Cid11} it is possible to determine the nature of the activity of NELGs that have weak or
undetected emission lines based on a new diagnostic diagram, the WHAN diagnostic diagram, that compares the EW of
H$\alpha$ with the ratio [NII]$\lambda6584/$H$\alpha$. Applying this diagram for a large sample of these galaxies taken
from SDSS, these authors concluded that most of them are ``retired galaxies'', that is galaxies where the activity is
dominated by PAGBs.

In Fig.~2 we show the WHAN diagrams for our three samples of LLAGN candidates. In Fig.~2a the LLAGN$\beta$ sample seems
to form a mixture of SFGs, Sy2s and LINERs. In Fig. 2b, the LLAGN$\gamma$ sample shows even more diversity: a small
number look like a mixture of SFGs and Sy2s, a higher fraction look like LINERs, and only a few look like retired
galaxies. In Fig.~2c the LLAGN$\alpha$ sample seems to be mostly composed of LINERs. Consequently, there seem to be very
few LLAGNs in our sample consistent with the PAGB hypothesis.  

From the above results, one could worry that the definition of retired galaxies is sensitive to noise introduced by the choice of a too low S/N selection criterion \citep[e.g.][]{RP94}. To test this hypothesis, we have changed our criterion, keeping only the NELGs with S/N $\ge 5$. The number of LLAGN$\gamma$ decreases to 82\% of the original sample with S/N $\ge 3$, the number of LLAGN$\beta$ to 77.0\% and the number of LLAGN$\alpha$ to 74.0\%.  However, in the WHAN diagram for the LLAGNs with S/N$\ge 5$ shown in Fig.~3, we can still identify a good fraction of LLAGN$\gamma$ as retired galaxies. Also, our description in terms of mixture of activity types stays the same. This result confirms that the reason why some emission lines are not detected in these galaxies is not due to low S/N.  For the rest of our analysis we will thus keep our original sample with S/N$\ge 3$.

For comparison sake, we trace in Fig.~4 the WHAN diagram for the standard NELGs classified as Sy2s and LINERs. Here we
can observe that although in the BPT-VO diagram we can discriminate between the LINERs and the Sy2s, the two classes
strongly overlap in the WHAN diagram. The difference comes from the use in the WHAN diagram of the EW, which is
sensitive to the stellar populations and morphologies of the galaxies, but not to the ionization state of the gas or the
nature of their ionizing source, which is related to the ratio of the emission lines [NII]$\lambda6584/{\rm H}\alpha$.
According to the standard definition, Sy2s are AGNs (they show high [NII]$\lambda6584/{\rm H}\alpha$ ratios) with high
ionization states, because they have high [OIII]$\lambda$5007/H$\beta$ ratios, while LINERs are AGNs with low ionization
states. This difference is not reflected in the WHAN diagram.

The situation is slightly better in Fig.~5 for the SFGs, but this is because the hosts of the SFGs form a more
homogeneous population, they are mostly late-type spirals. However, we also note an important overlap with the Sy2s,
which does not exist in the BPT-VO diagram. The situation becomes even more ambiguous in the case of the TOs. However,
according to the WHAN diagram most of these galaxies are AGNs, classified either as Sy2s or LINERs, not SFGs.

According to the WHAN diagram, it is difficult to understand why the AGN nature of our three groups of LLAGN candidates was
not recognized. This is unless one also rejects the AGN nature of LINERs \citep[e.g.][]{EHF10,Singh13}. So it seems
fundamental in our analysis that we also verify the PAGB hypothesis for the LINERs (see Section~4.1).

\subsection{Covering the entire post-AGB evolutionary phase}

To be able to test the PAGB hypothesis, it is important to build a sample of stars that
covers all the different evolutionary stages of the post-AGBs: the genuine PAGBs
(pre-PNs), the PNs and the hot white dwarfs.

For the PAGBs we adopted the definition of \citet{Sua06} and used their list. Using a search radius of 5 arcseconds,  we
cross-correlated the positions of these stars with those of the objects in the WISE
catalog and kept only the stars with fluxes that have $S/N > 2$ in all the four MIR wavebands.
The spectra of these PAGBs \citep{Sua06} show no emission lines, confirming that they
are pre-PNs. From an original sample of 101 PAGBs, we found 76 stars satisfying our
selection criteria ($\sim76$\% of the whole sample). 

For the PN phase, we used the list of 492 confirmed PNs as published by \citet{WG11}.
Applying the same search and quality of flux criteria in WISE as for the PAGBs, we found
419 PNs ($\sim85$\% of the whole sample).

To complete the AGB evolutionary phase, we used the new catalog of spectroscopically
confirmed white dwarf stars from SDSS DR 7 as published by \citet{Kleinman13}. There are
20407 white dwarfs in this catalog, 30\% of which are hot (with an effective temperature
above or equal to 18000~K). Applying the same search criteria in WISE as for the
PAGBs and PNs, we found only 4056 stars ($\sim 20\%$ of the whole sample). Furthermore, all, except a
few, only have upper limits in 12 and 22$\micron$ bands. Among these 4056 ``partially detected'' stars, we count 1370 hot
white dwarfs ($\sim33$\% of the sample), which implies that the ``non-detection'' of white dwarfs at 12 and 22$\micron$ in
WISE is independent of the effective temperature of the stars.

Our search results for the white dwarfs suggests that many of these stars are either
completely free of dust, or that collectively their light is too diluted to be able to
heat the dust over kpc regions in our galaxy. However, this does not mean that we cannot test for the
presence of hot white dwarfs in LLAGNs using MIR observations (see our discussion in Section~4).

\section{Results: the MIR colors of LLAGNs}

In Fig.~6 we present the color-color diagram [4.6]-[12] vs.~[4.6]-[22] for each of the
different LLAGN candidate samples, and compare their MIR colors with those of the PAGBs and
PNs. For the stars we show the individual colors whereas for the galaxies we trace the
distribution of their colors as normalized density contours, including the
corresponding fraction (25\%, 50\% and 75\%) of galaxies in the sample around the point of maximum density (the
most probable colors). The differences between the stars and galaxies in these color-color
diagrams are compelling. Contrary to what is expected based on the PAGB hypothesis, only a few
PAGBs and PNs have colors consistent with those of the LLAGNs. In general, the PAGBs and
PNs are much redder than the LLAGNs. Note that other combinations of MIR colors yield similar
results, with those of the stars significantly different than those of the LLAGNs.

\begin{figure}[!t]
\begin{center}
\includegraphics[width=0.8\textwidth]{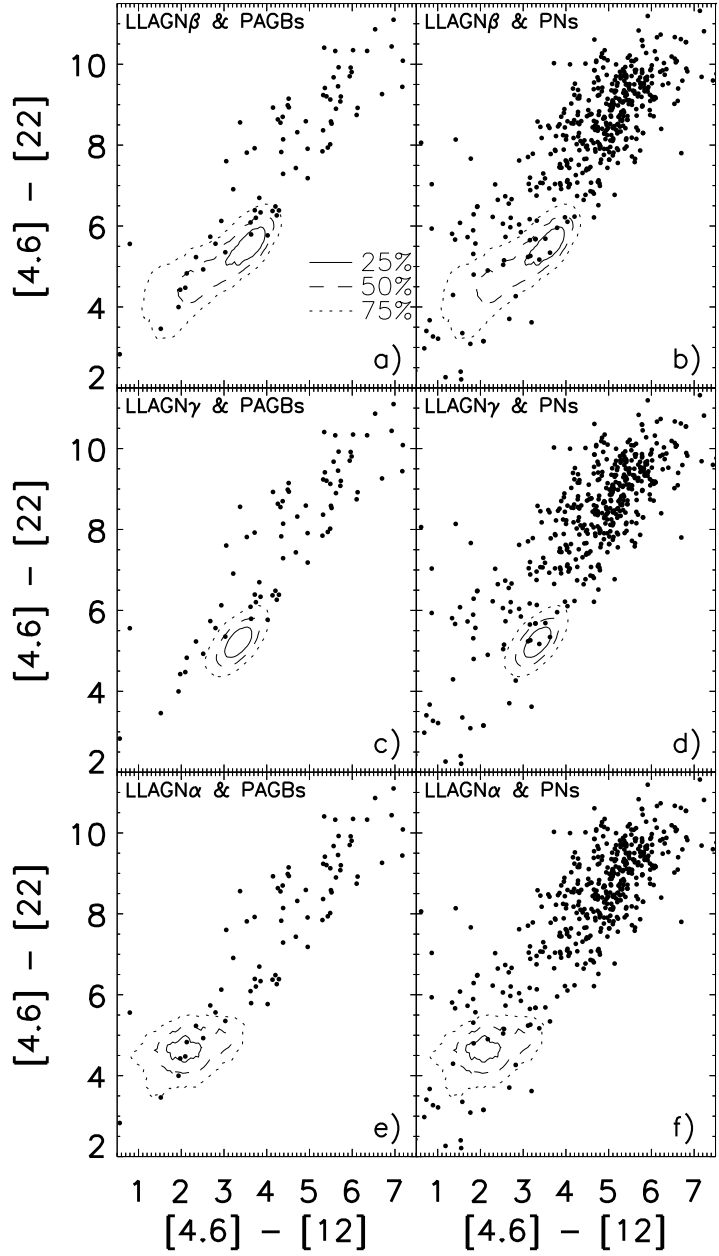}
\caption{MIR color-color diagrams formed by comparing the differences in magnitude between
the 4.6 and 12 micron bands, with the differences in magnitude between the 4.6 and 22
micron bands. In each panel we draw the density contours (normalized to 100\% at
the peak density)
for the galaxies, while the dots are the colors of the individual stars. }
\label{fig:6}
\end{center}
\end{figure}

\begin{figure}[!t]
\begin{center}
\includegraphics[width=0.8\textwidth]{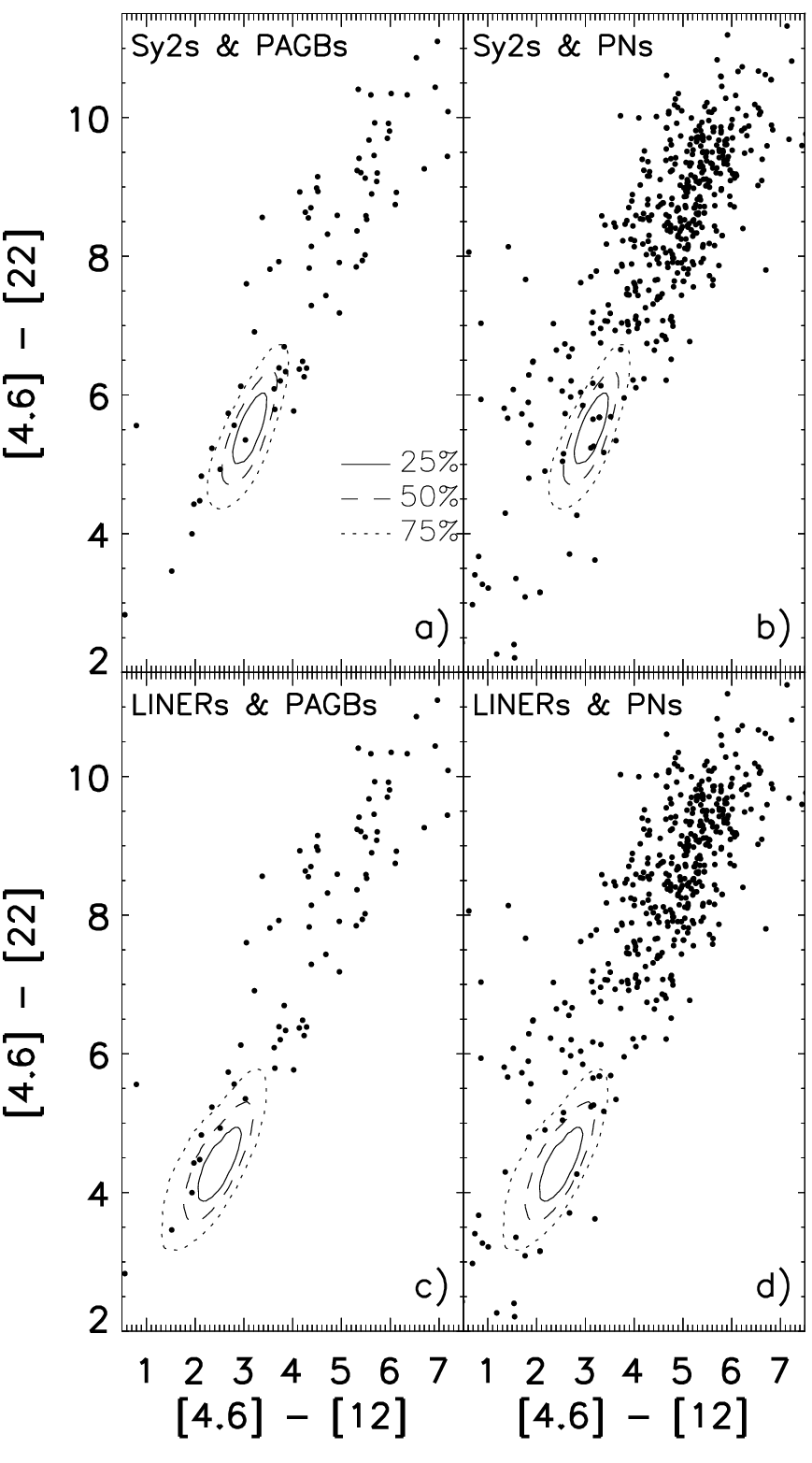}
\caption{Same as Fig.\ 6 for the Sy2s and LINERs in the standard NELGs sample.}
\label{fig:7}
\end{center}
\end{figure}

\begin{figure}[!t]
\begin{center}
\includegraphics[width=0.8\textwidth]{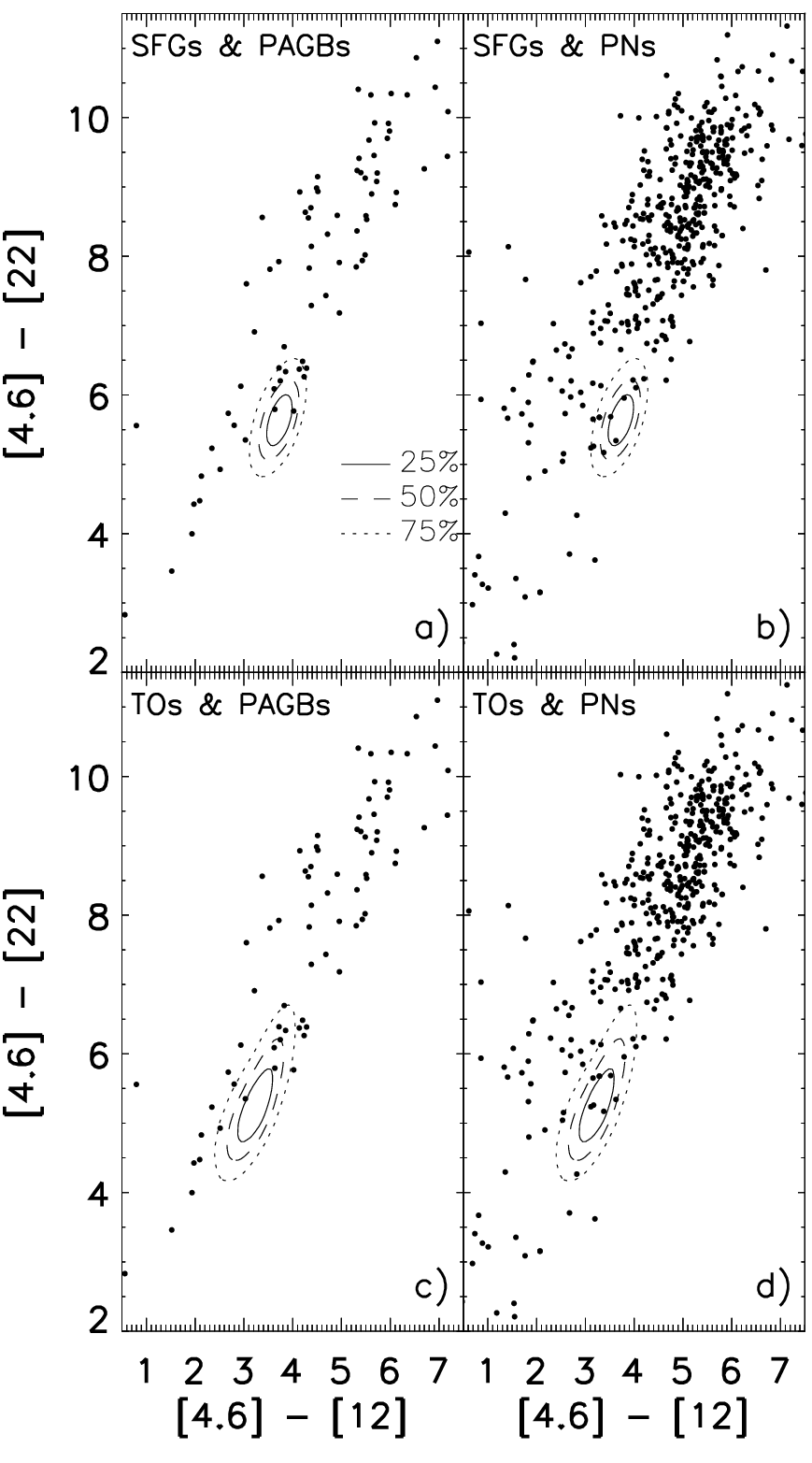}
\caption{Same as Fig.\ 6 for the SFGs and TOs in the standard NELGs sample.}
\label{fig:8}
\end{center}
\end{figure}

In general, the colors for the stars differ from those of the LLAGNs by two important
characteristics: their dispersions are much larger and their magnitudes show greater
variations passing from one band to the other. This second characteristic for the stars is
an indication that their MIR continua trace SEDs than are different from those of the LLAGNs (see our
discussion below).

We show similar color-color diagrams for the Sy2 and LINERs in Fig.~7 and for the SFGs and
TOs in Fig.~8. All the standard NELGs in our large sample show well defined colors, which
is contrary to what is observed for the PAGBs and PNs. Although the colors of the standard
NELGs are redder than those of the LLAGNs, the differences between the stars and the
galaxies are still too large to be reconciled with the PAGB hypothesis. Again, using other
combinations of colors yield similar results.

In Fig.\ 9 we compare the box-whisker plots for the MIR colors of the PAGBs and PNs with
those of the NELGs. The NELGs become bluer following the sequence SFGs $\rightarrow$
TOs-Sy2s-LLAGN$\beta$ $\rightarrow$ LLAGN$\gamma$ $\rightarrow$ LLAGN$\alpha$-LINERs. The colors of the
LLAGNs and LINERs are extremely blue compared to the colors of the PAGBs and PNs. The
sequence in colors observed is in direct contradiction with what is expected based on the PAGB hypothesis.

\begin{figure}[!t]
\begin{center}
\includegraphics[width=0.65\textwidth]{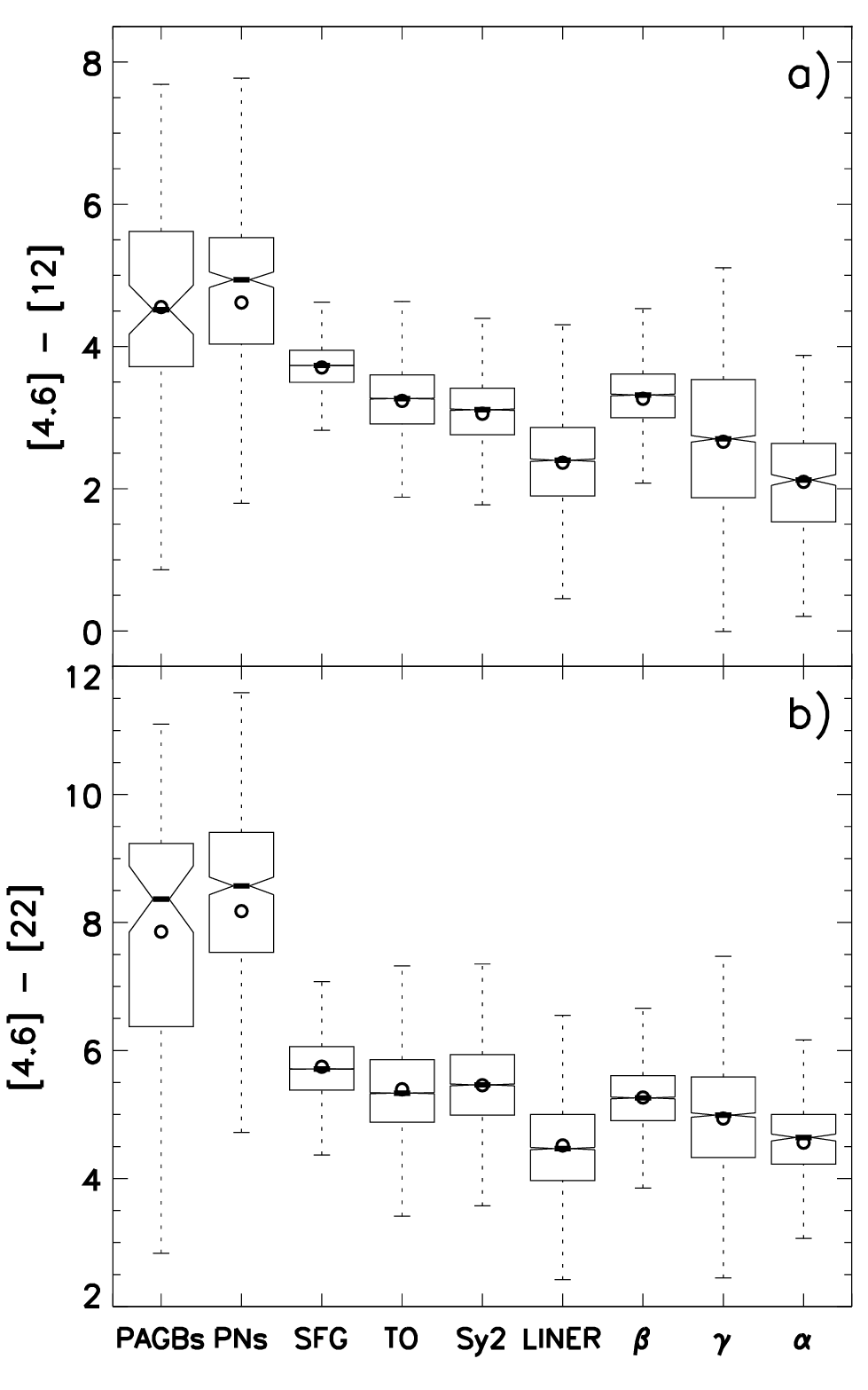}
\caption{Box-whisker plots comparing the two MIR colors used in our study as observed in the
PAGBs, the PNs, the standard NELGs and the three LLAGN candidates, which are identified by their greek letters. 
The lower side of the box is the lower quartile, $Q_1$, and the upper side is the upper quartile,
$Q_3$. The whiskers correspond to $Q_1 - 1.5\times IQR$ and $Q_3+ 1.5\times IQR$, where $IQR$ is the interquartile range. The
median is shown as a bar and the mean as a circle. The notches have a width proportional
to the $IQR$ and are inversely proportional to the square root of the
size of the sample $N$. Comparing two samples, no overlapping notches implies that the medians of
the two samples have a high probability of being different.}
\label{fig:9}
\end{center}
\end{figure}

\begin{figure}
\includegraphics[width=12cm,height=9.5cm]{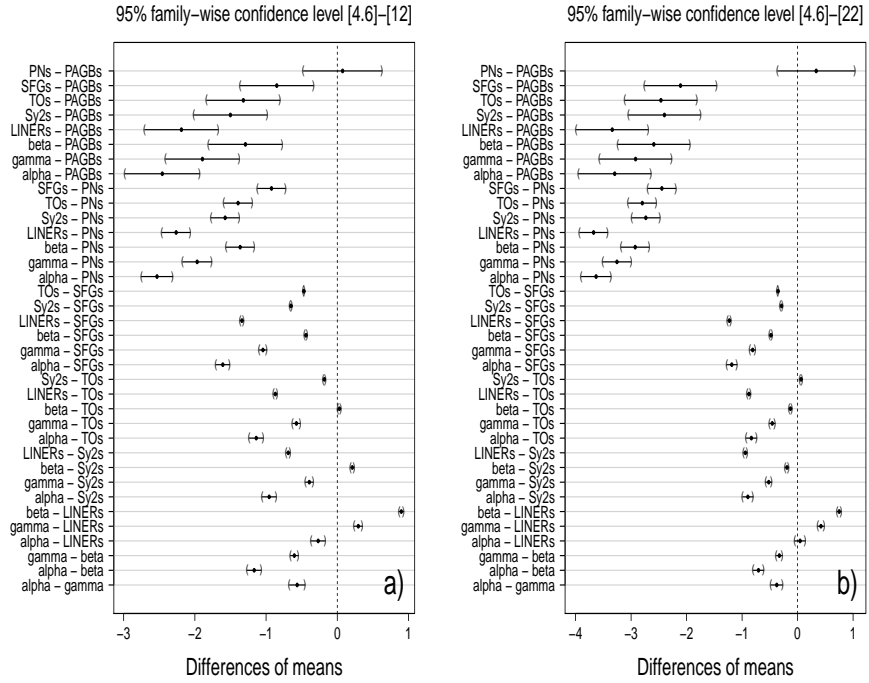}
\caption{The simultaneous 95\% confidence intervals for all pairwise comparisons of the
subsample means in colors ($\bar{C_i} - \bar{C_j}$) as obtained using the max-t test: a) for the [4.6]-[12] colors, b) for the [4.6]-[22] colors. A confidence interval including zero indicates no statistically significant
difference in colors, and the farther from zero the more significant the difference.
Negative differences suggest the objects in the subsample $i$\ are bluer on average than
the objects in the subsample $j$. Comparing samples with small dispersions (small $IQR$)
and large sample size produce small confidence intervals, making the results of the test
more significant.}
\label{fig:10}
\end{figure}

To test the statistical significance of the differences in
colors, we present in Fig.~10 the simultaneous 95\% confidence
intervals for all the pairwise comparisons of the subsample color
means. These confidence intervals were obtained using the max-t test,
which is based on a parametric ANOVA model \citep{HBW08,HSH10} that
does not suppose the variances are the same (heteroscedasticity)
or that the sizes of the samples are comparable. This test assumes
the null hypothesis takes a linear form and estimates the confidence
interval calculating the difference between the means in color, adding
and subtracting (upper limit and lower limit) the standard error,
then multiplying by the value of a t-distribution, adopting a level
of confidence of 95\%. A confidence interval including zero indicates
no statistically significant difference between the subsample means,
and the farther from zero the more significant the difference.

According to the results of the max-t test presented in Fig.~10 there is no significant
difference on average between the colors of the PAGBs and PNs. A similar result was
obtained before by \citet{Sua06} in the far infrared using IRAS. It is obvious from the
max-t test that the NELGs, and most particularly the LINERs and LLAGNs, are too blue (all
the differences are negative) to be compatible with the dominant presence of PAGBs or PNs
in any of these galaxies. Based on these results, we can confidently reject the PAGB
hypothesis for the LLAGNs in our sample.

\begin{figure}[!t]
\begin{center}
\includegraphics[width=0.7\textwidth]{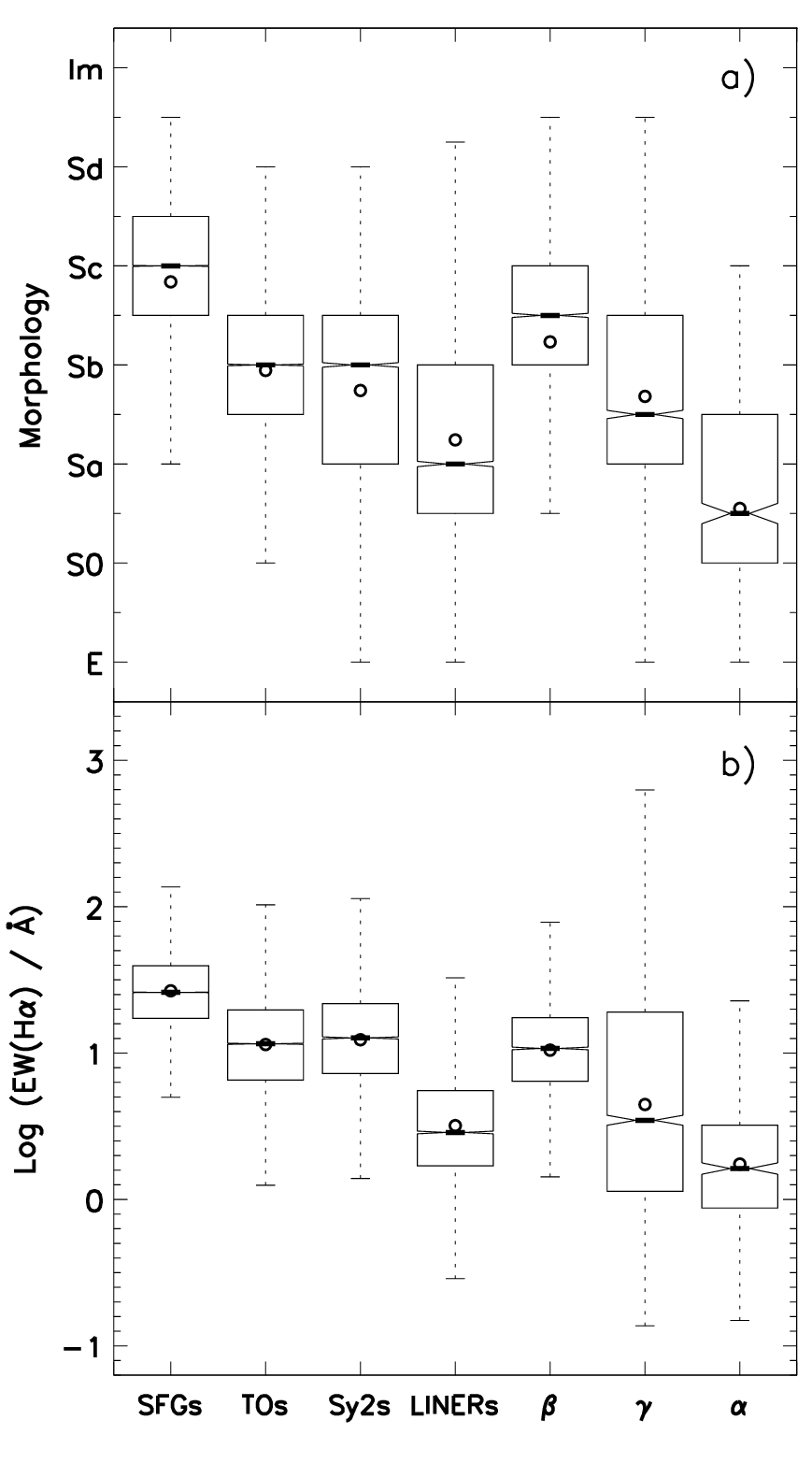}
\caption{Box-whisker plots comparing (a) the morphologies of the 
NELGs in our sample, and (b) their H$\alpha$ EWs. The explanations for the box-whisker 
plots and identification of the samples are as in Fig.~9.}
\label{fig:11}
\end{center}
\end{figure} 

The max-t tests also confirm the color sequence seen in Fig.~9: the NELGs become bluer following the
sequence SFGs $\rightarrow$ TOs-Sy2s $\rightarrow$ LLAGN$\beta$ $\rightarrow$
LLAGN$\gamma$ $\rightarrow$ LLAGN$\alpha$-LINERs. In \citet{TP13} a similar sequence was
found for the standard NELGs, which was shown to trace a gradual decrease in star formation activity as the
morphologies of the galaxies change from late to early types \citep{Ken92a,Ken92b,Coz96,Coz11}.
This is supported by Fig.~11, where we compare the box-whisker plots for the morphologies
and EWs of H$\alpha$\ for all the NELGs in our sample, separated by activity types. The
morphologies of the standard NELGs were determined in \citet{TP12a} and the same method was
used to determine the morphologies of the three different samples of LLAGN candidates in
this study. One can see that the EWs of H$\alpha$ for the standard NELGs decrease
following the sequence SFGs $\rightarrow$ Sy2s-TOs $\rightarrow$ LINERs. Following the
same sequence these galaxies pass from late-type spirals to early-type spirals. Similarly,
the EWs for the three LLAGN samples decrease following the sequence LLAGN$\beta$
$\rightarrow$ LLAGN$\gamma$ $\rightarrow$ LLAGN$\alpha$ and following the same sequence
the morphologies change from late-type spirals to early-type spirals.

The above results suggest a simple physical explanation for the non-detection of some
emission lines in the LLAGNs. The high ratios [NII]$\lambda6584/{\rm H}\alpha$ of the
LLAGNs in the WHAN diagram suggest that these galaxies are mostly metal rich \citep[as was
previously recognized also by][]{Cid10}. If we could trace the BPT-VO diagram for these
galaxies, most probably they would  be located at the junction between the SFGs and AGNs
arms of the $\nu$ shape distributions, since this is where we find the most metal rich
NELGs \citep{Coz11,TP12b}. Therefore, the intensity of the [O~III]$\lambda$5007 emission
line is expected to be lower than the intensity of the H$\beta$ emission line by a factor
3, 10 or more. This characteristic is at the basis of the explanation for the non-detection
of some emission lines in NELGs. 

According to the color sequence, the LLAGN$\beta$ are
still actively forming stars, which suggests that their spectra are surely affected
by dust absorption. Both emission lines in the blue, [OIII] and H$\beta$, would thus
decrease in intensity. However, considering the high intensity difference, [OIII] could
easily disappear while H$\beta$ would still be observed. On the other hand, the color
sequence for the LLAGN$\gamma$ suggests that the star formation in these galaxies as compared to the LLAGN$\beta$ has already
started to decline, which implies that these galaxies
may be particularly rich in intermediate age stellar populations \citep{DW86,Coz96,CDD01}. Since the
Balmer absorption lines are culminating in intermediate age stars \citep{Rose85}, the
H$\beta$ emission line could completely disappear. At the same time, if the LLAGN$\gamma$
have slightly lower metallicity than the LLAGN$\beta$, then the intensity of the [OIII]
lines could be sufficiently high to be detected. Finally, the color sequence for the
LLAGN$\alpha$ suggests star formation, like in the LINERs, is at its lowest level in these
galaxies. Then, in high metallicity galaxies both emission lines could easily become
undetected.

We conclude therefore that a variation in star formation \citep{TP13}, in parallel with
variations in morphology and metallicity explain the MIR colors sequence and the
differences between the different standard NELGs and LLAGNs. According to this
explanation, about 20\% of the NELGs have undetected lines, because they are especially
metal rich, and, as we observed, this is independent of the S/N in their spectra.

\section{Discussion}

\subsection{Revealing the AGN nature of LLAGNs}

In the previous section we have shown that the MIR colors of the LLAGNs are significantly different from those of the
PAGBs and PNs. Only a few of these stars have colors consistent with those of the LINERs and LLAGNs in our sample.
However, for the PAGB hypothesis to be satisfied, the PAGBs and PNs must define the colors of the LLAGNs, which implies
that their color distributions must be exactly the same. Also, considering the high number (of the order of $10^{11}$) of hot
white dwarfs required to explain the ionization of the gas in a typical LLAGN, it is obviously not sufficient to have a few "consistent" cases, as it would imply an even higher number (higher by at least a factor 10) of special PAGBs and PNs is
needed. Moreover, the few consistent cases observed are not specific to the LLAGNs or LINERs, since we can also find
PAGBs and PNs that fit the Sy2s, the SFGs and the TOs. We conclude that the PAGB hypothesis cannot be maintained on the basis of
the data we presented.

\begin{figure}[!t]
\begin{center}
\includegraphics[width=0.7\textwidth]{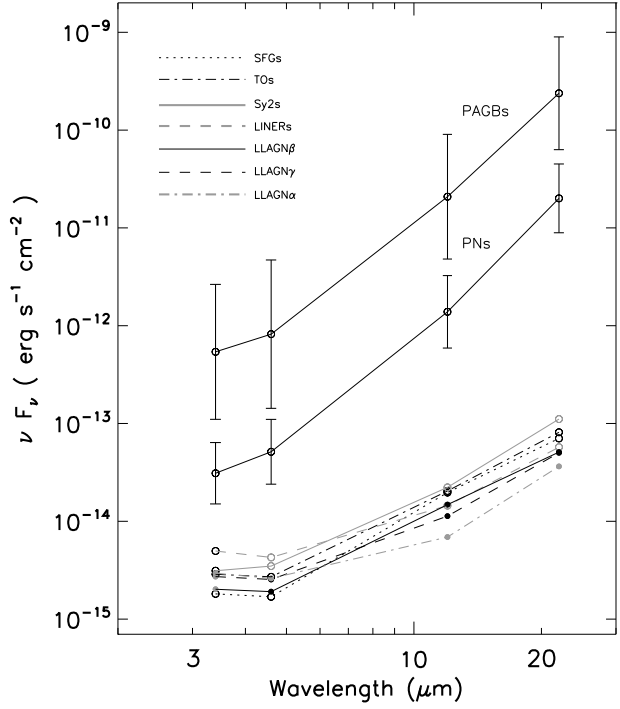}
\caption{The mean MIR fluxes at the four different wavelengths used by WISE. How one transforms the WISE magnitudes into fluxes is explained in \citet{Wright10}. The
error bars for the stars are the errors on the means \citep{BR03}. For clarity sake, the errors
on the mean for the galaxies are shown only in Fig.~12. }
\label{fig:12}
\end{center}
\end{figure} 

\begin{figure}[!t]
\begin{center}
\includegraphics[width=0.7\textwidth]{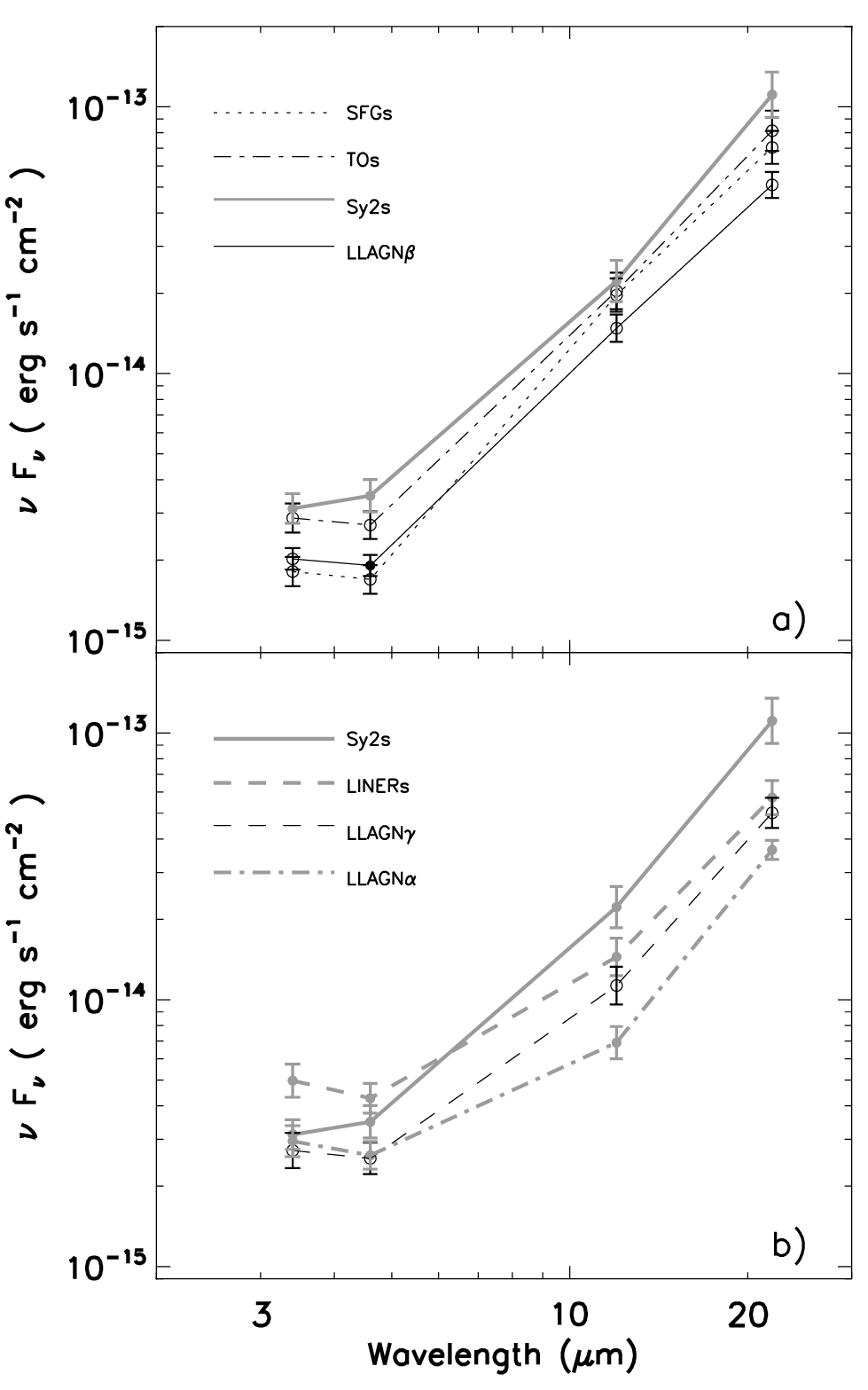}
\caption{Same as in Fig. 12, showing in a)  
the NELGs with evidence of recent star
forming activity and in b) the NELGs with no
such evidence. The data for the Sy2s were repeated in b for comparison sake.}
\label{fig:13}
\end{center}
\end{figure}

The difference in MIR colors between the stars and the NELGs in our sample
suggests that their continua trace different types of SEDs. In Fig.~12 we show the mean MIR fluxes in each band as a function
of the wavelength. We see that in both the PAGBs and PNs the fluxes vary greatly as we go to higher wavelengths, suggesting that
the SEDs of the PAGBs and PNs have steep slopes in the MIR. This is consistent with black bodies with high
temperatures \citep[see Figure~1 in][]{Vickers14}. On the other hand, the fluxes show much smaller variations in the NELGs, suggesting that their SEDs in the MIR are flatter than those of the stars. For the SFGs, this is consistent with
black bodies that have lower temperatures than those fitting the PAGBs and PNs \citep[][]{Anderson12}. For the other NELGs, however, 
still flatter SEDs are needed, like those typically found in AGNs \citep[e.g.][]{Polletta07,Yan13}.

The MIR data are consistent with a simple model where the SEDs of the NELGs are formed by a mixture of black bodies with a low
temperature typical of dust heated by O and B stars in HII regions, and power laws due to AGNs, which
have flatter MIR continua than the SFGs. The relative importance of the two continuum components in NELGs is better
observed in Fig. 13. According to \citet{TP13} we have separated the NELGs in two groups: in a) we show the mean fluxes
 in the NELGs that present a high level in recent star formation (the SFGs,
TOs and Sy2s) and in b) we show the flux variations in the LINERs, where star formation is presently at a very low level
(the data for the Sy2s were included for comparison sake). One can see that the the variation of flux with wavelength becomes larger as the
level of star formation in the galaxies increases, i.e. the SED steepens as the level of star formation in the galaxies increases. Based on the flux
variations the LLAGN$\beta$ are similar to galaxies in the high star formation group, while the LLAGN$\gamma$ and
LLAGN$\alpha$ are similar to galaxies in the low star formation group. This is consistent with the classification
obtained with the WHAN diagram, where it was found that the LLAGNs form a mixture of SFGs and AGNs. This result suggests that the continua vary with the amount of star formation, becoming flatter in
galaxies where the contribution of an AGN becomes predominant \citep[][]{Mat12,Donoso12,Jarrett13,Rosario13}.

It is easy to understand why a power law explains the trend for the colors to become bluer in the LINERs and LLAGNs.
For example, consider a power law $L_\nu = C \nu^{-\alpha}$ with an exponent $\alpha = 0$, which yields fluxes
independent of the wavelength. Such continuum would produce comparable MIR fluxes, and, consequently, smaller
differences in magnitudes, which is consistent with the blue color trend \citep[c.f. compare the colors of the power
laws and black bodies in Table~1 of][]{Wright10}. Therefore, we argue that due to a decreasing level of star formation
in the LLAGNs and the LINERs, the continua become flatter and the MIR colors bluer, which we interpret as evidence that the power-law components due to the AGNs heating the dust in these galaxies become more apparent.

 \begin{figure}
\includegraphics[width=0.85\textwidth]{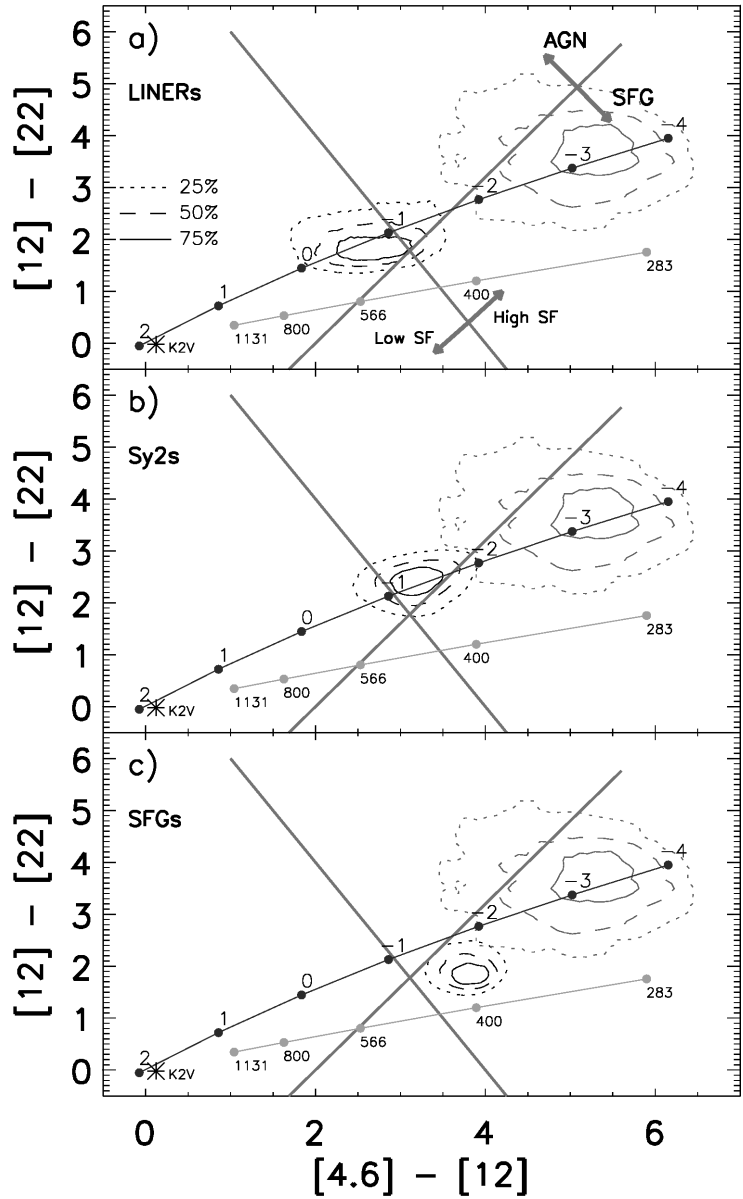}
\caption{New diagnostic diagram using the WISE colors. The solid line with a positive slope separates AGNs from SFGs. The solid line with a negative slope separates galaxies with high star formation activity from galaxies with low star formation activity (see explanations in the text). The positions of the LINERs, the Sy2s and SFGs are given in terms of normalized density contours, centered on the most probable colors for the activity types. For comparison sake, the density contours for the PAGBs and PNs samples added together are repeated in the three diagrams. Drawn over the data we also show the colors produced by a power law with different exponents, and a black body with different temperatures \citep{Wright10}. As a point of reference, a star indicates the position occupied by a typical K2V star. }
\label{fig:14}
\end{figure}

Based on the analysis described above, it should be possible to separate the NELGs according to their dominant continuum component. To verify this assumption, we present in Fig.~14 the color-color diagrams [4.6]-[12]
vs.~[12]-[22] for the LINERs, the Sy2s and SFGs. We have also traced over the data the estimations made by
\citet[][]{Wright10} of the colors expected in WISE for a power law with different exponents, and a black body at
different temperatures. In Fig.~14 one can see that the LINERs and Sy2s have colors consistent with a power law, with an
exponent $\alpha$ varying between 0 and $-1.5$\ in the LINERs and between $-1$\ and $-2$\ in the Sy2s. The SFGs on the
other hand show colors that are intermediate between those produced by a power law or a black body. The variation of the
[4.6]-[12] color in the NELGs with different activity types suggests that this color is sensitive to the present level
of star formation in these galaxies \citep[for similar conclusions see][]{Mat12,Rosario13}.

In Fig.~14 we have also included the color distribution for the PAGBs and PNs samples added together. Surprisingly their colors also seem to be fitted by a power law. This is probably due to the fact that the SEDs of these stars must be fitted by multiple black bodies with different temperatures \citep[see][]{Anderson12,Vickers14}, the sum of which mimics a power law distribution. However, the power law for the PAGBs is much steeper than for the LINERs or Sy2s, the exponent $\alpha$ varying between $-2$\ and $-4$.  Consequently, the PAGBs and PNs can clearly be distinguished from the NELGs in this diagram. 

According to our analysis, there is no evidence in MIR for a high number of PAGBs or PNs in the LINERs. This goes against the PAGB hypothesis, which claims that PAGBs are responsible for ionizing the gas in these galaxies. However, what can we say about the hot white dwarfs? In our galaxy these stars turned out to be undetectable in the 12 and 22 $\micron$\ bands, which suggests that their SEDs must be extremely blue, falling abruptly in the MIR \citep[e.g.][]{Blommaert06}. The PAGB hypothesis would thus lead to a new dilemma: the source of the ionizing photons in LINERs must be different from the source of photons that heat the dust. But, from our analysis of the MIR colors, we saw that an AGN can easily explain the colors of the LINERs, and if an AGN exists in these galaxies, then this AGN can also be the source of the ionizing photons, which makes the PAGB hypothesis redundant.   

Alternatively, one can suggest, without direct observational evidence, that the hot white dwarfs somehow are also responsible for heating the dust in the LINERs. After all, some PAGBs and PNs in our analysis do reproduce the MIR colors of some of the NELGs in our sample (note, however, that the colors are not specific to one activity type) and one could propose to identify the SEDs of these special cases to the typical SEDS of hot white dwarfs in LINERs. However, there are many problems with this hypothesis. For example, reproducing the flat SEDs required to explain the LINERs in our sample implies black bodies that have lower temperatures than those of O and B stars in the SFGs. But the huge amount of ionizing photons necessary to explain the ionized gas in a typical LLAGN constitutes a strong constraint on the temperature of the stars, and the temperature cannot be lowered arbitrarily without increasing significantly the number of hot white dwarfs. Alternatively, one may assume very different distributions for the gas and the dust, namely large dust free ionized regions bounded by huge clouds of dust located at distances farther from the stars than the gas (the dust is not mixed homogeneously with the gas). But the relatively strong emission observed in MIR constrains the quantity of dust required, and the amount of dust cannot be increased arbitrarily either. Also, according to the current knowledge about the SEDs of PAGBs and PNs in the MIR \citep[e.g.][]{Anderson12,Vickers14} and according to our own analysis, it seems difficult to contrive a simple scenario that would fit specifically the SEDs of the LLAGNs based on the SEDs of hot white dwarfs. To explain the MIR colors of the LINERs, therefore, the white dwarf hypothesis seems more complicated, and consequenlty less credible than the AGN model.

\subsection{Defining a new diagnostic diagram in MIR}

\begin{figure}
\includegraphics[width=0.9\textwidth]{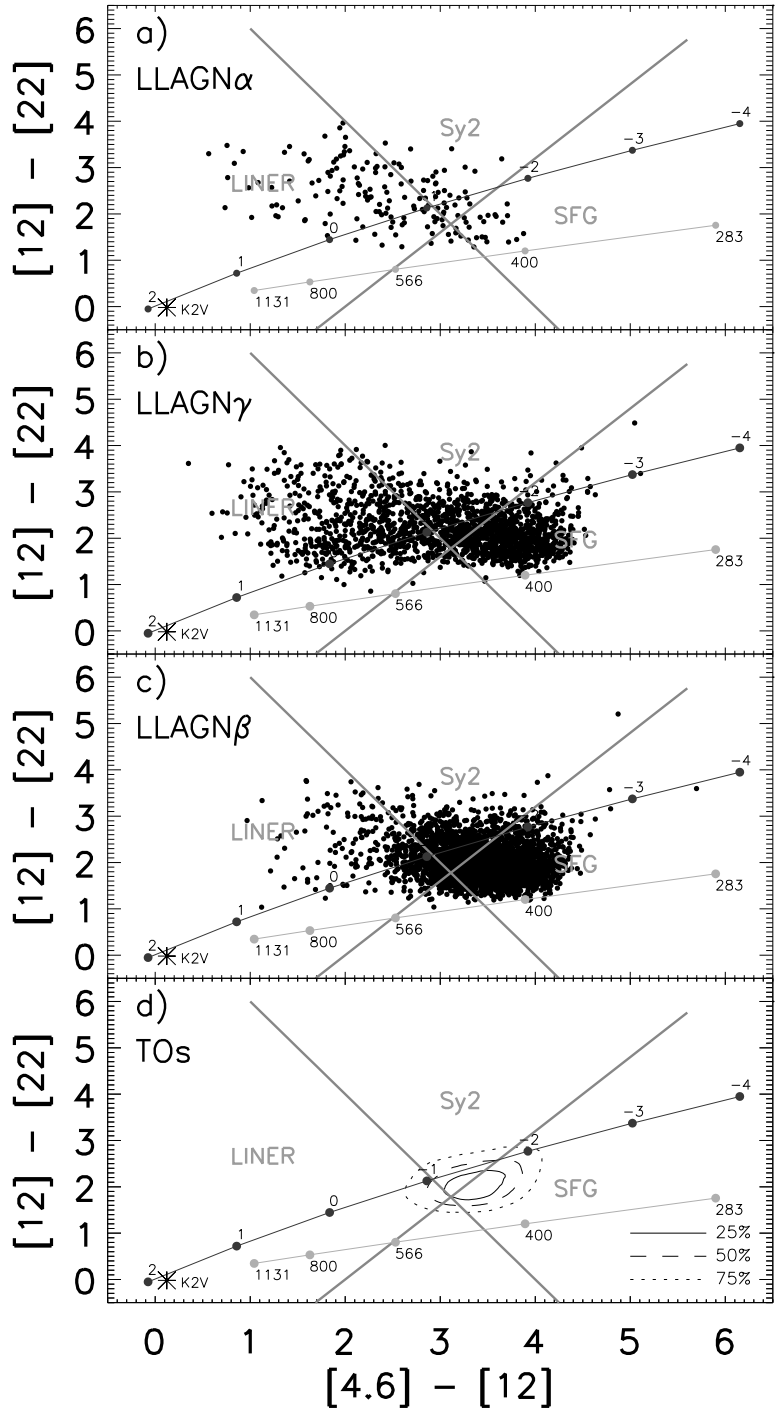}
\caption{The MIR diagnostic diagram for a) the LLAGN$\alpha$, b) the LLAGN$\gamma$, c) the LLAGN$\beta$, and d) the TOs.}
\label{fig:15}
\end{figure}

We can use the color-color diagram in Fig.~14 as a new diagnostic diagram. The physical
justification for the two empirical separations shown in this diagram is the following: the NELGs with different activity types can be distinguished based on their dominant continuum component and based on their present level of star formation \citep{TP13,Rosario13}. Consequently, we have traced in Fig.~14 a diagonal
(with a positive slope) that separates the SFGs from the AGNs, and another diagonal (with a negative slope) that
separates the AGNs with low star formation activity (LINERs) from the AGNs with high star
formation activity (Sy2s). Note
that to define the different activity zones we have used only the MIR fluxes with S/N $ \ge
3$ (corresponding to quality flags, ph\_qual, equal to A or B, in all of the four WISE
bands). The density contours suggest that one can only obtain a probability of about 50\%
percent for the separation between LINERs and Sy2s. However, according to this diagram the
probability that a SFG could be mistaken as an AGN is smaller than 25\%. In fact, the
probability that a SFG could be mistaken as an AGN is zero, if one considers the different [NII]$\lambda6584/$H$\alpha$ ratios of
these galaxies (c.f. Fig.~1). 

In Fig.~15 we show the classification obtained for the LLAGN candidates and the TOs using the new MIR diagnostic diagram. It
can be seen in Fig.~15a that the LLAGN$\alpha$ are similar to the LINERs. This suggests that they are AGNs with a low
level of star formation activity. The LLAGN$\gamma$\ (Fig.~15b), on the other hand, form a mixture of SFGs, Sy2s and
LINERs, which again is consistent with our previous analysis. It is interesting to compare the LLAGN$\beta$ in Fig.~15c,
with the TOs in Fig.~15d. The MIR colors of the TOs suggest they are intermediate between the SFGs and AGNs. This is
consistent with the standard interpretation based on the BPT-VO diagram \citep{Kew01,Kew06}. The LLAGN$\beta$ would thus
be similar to the TOs, which are LLAGNs with a high level of star formation. 

Considering the color sequence traced by the LLAGNs in Fig.~15, we see that as the [4.6]-[12] color decreases, the colors of the LLAGNs become similar to those of the LINERs, i.e. there is a clear change of colors from those produced by black bodies in SFGs to those consistent with power laws in AGNs. Because the [4.6]-[12] color is sensitive to the present level of star formation, we believe 
highly probable that as the star formation decreases in the LLAGNs, their MIR colors will change for those of the
LINERs \citep{Lee07,WW08,Chen09,Chen10,Yuan10,TP12a,Zhang13,TP13}. Consequenlty, we conclude that what distinguishes the
different LLAGNs in our sample is their different levels of star formation. When the star formation activity is high, the
LLAGNs look like TOs, and when the star formation is declining they look like Sy2s or LINERs, which are all AGNs. 
Therefore, the PAGB hypothesis is also redundant in the case of the LLAGNs. 

In general, the results obtained using the new MIR diagnostic diagram are consistent with those obtained with the BPT-VO
diagram for the standard NELGs and the WHAN diagram for the LLAGNs. However, according to our analysis there are no
more ambiguities as to the AGN nature of the LINERs, and consequently for the other LLAGNs, since there are only two main sources for heating the dust, namely massive stars and AGNs
\citep{Wright10,Jarrett11,Mat13,Assef13}. The MIR diagnostic diagram also suggests that these two mechanisms are frequently working at the
same time \citep{HFS97,Wu98,Maoz99,VV00,Kew01,Carter01,CZ06,GD08,Yuan10,Carpineti12}. Indeed, what distinguishes the LINERs, the Sy2s and SFGs and explains the distributions of the LLAGNs in the MIR diagnostic diagram is their different levels of star formation 
\citep{Lee07,WW08,Chen09,Chen10,Yuan10,TP12a,Zhang13,TP13}. 

Another
advantage of this new MIR diagnostic diagram is that it should allow to classify different types of galaxies using the same
criteria. This includes galaxies that cannot be classified using standard diagnostic diagrams, like the LLAGNs, but also radio galaxies and
broad-line AGNs, like the Sy1s and the quasars.

\section{Summary and conclusions} \label{sec:conclusion}

For many authors in the field, LLAGNs do not exist. They are retired galaxies, misidentified as AGNs, where the gas is
ionized by PAGB stars. However, according to the PAGB hypothesis, of the order of $10^{11}$ hot white dwarfs would be
required to produce the right amount of ionizing photons in a typical LLAGN. Such a high number of hot stars must leave
a trace not only in the optical but also in the infrared. Indeed, using MIR observations obtained with WISE we have found that an important
fraction of LLAGNs are emitting in infrared and that their colors are characteristically blue. However, by comparing the
MIR colors of the LLAGNs with those of the PAGBs and PNs in our galaxy, we have found that only a
few of these stars have colors similar to the LLAGNs, the majority being extremely red, which contradicts the PAGB hypothesis. 

The main reason why the PAGBs colors differ from those of the LLAGNs is because they have different SEDs. The SEDs of
the stars in the MIR have steep slopes, while those of the LLAGNs are flatter. It is the flatness of the SEDs that
produces the blue colors. These differences can be explained by assuming a black body with high temperatures in the PAGBs
and a mixture in the LLAGNs of black bodies at low temperature, consistent with dust heated by O and B stars in HII regions, and
different power laws due to AGNs, which have flatter MIR continua than the SFGs. As the level of star
formation in the LLAGNs decreases, their continua become flatter, which we interpret as evidence that the power-law
components due to the AGNs heating the dust in these galaxies become more apparent.

Consistent with the AGN model, we have shown that in a color-color diagram the LINERs and Sy2s have colors consistent
with a power law with an exponent $\alpha$ varying between 0 and $-1.5$\ in the LINERs and between $-1$\ and $-2$\ in
the Sy2s. Surprisingly we have also found the colors of the PAGBs to follow a power law, but with a much 
steeper exponent than for the AGNs. We have also shown that the variation of the [4.6]-[12] color allows to distinguish
between the NELGs with different activity types, which suggests that this color is sensitive to the present level of star
formation in these galaxies.

We conclude, therefore, that the MIR characteristics of the LINERs and of the other LLAGNs in our sample are consistent
with the view that they are genuine AGNs
\citep[][]{Hec80,OD83,Ho93,Barth98,Satyapal04,Sarzi05,Kew06,Filho06,CZ06,GM06,GM09,Kau09,Coz11,TP12a,TP13}. If an AGN is responsible for heating the dust in these galaxies, then this AGN may also be responsible for ionizing the gas, which makes
the PAGB hypothesis redundant. 

The LLAGNs show a large variation in their level of star formation, which explains their
different characteristics. However, the MIR color sequence suggests that as the star formation decreases in these
galaxies, they may all resemble the LINERs. Therefore, we also conclude that galaxies with a LINER spectral
characteristic are akin to evolved or dying quasars \citep{Hec80,Kau09,Coz98,Ric98,Mil03,Gav08}.

Thanks to our analysis we have found a new diagnostic diagram in the MIR that confirms the classification obtained
using the BPT-VO diagram by eliminating the ambiguity as to the AGN nature of the LINERs. One other important advantage
of this new diagnostic diagram is that it allows to determine the nature of the activity in any type of galaxies using
the same criteria. This includes galaxies that cannot be classified using standard diagnostic diagrams, like the LLAGNs
and the many radio galaxies, as well as broad-line AGNs like the Sy1s and the quasars. This diagnostic diagram represents a
powerful new addition to the panoply of tools already in use to study active galaxies.

\section*{Acknowledgments}
We thank an anonymous referee for making comments and suggestions that helped us improving the quality of our study and its
presentation. J.P. T.-P. acknowledges PROMEP for grant 103.5-10-4684, and DAIP for grant DAIP-Ugto (0065/11). This
research has made use of the VizieR catalogue access tool, CDS, Strasbourg, France:http://vizier.u-strasbg.fr/. This
publication made use of data products from the Wide-field Infrared Survey Explorer (WISE), which is a joint project of
the University of California, Los Angeles, and the Jet Propulsion Laboratory/California Institute of Technology, funded
by the National Aeronautics and Space Administration. The funding for Sloan Digital Sky Survey (SDSS) has been provided
by the Alfred P. Sloan Foundation, the Participating Institutions, the National Science Foundation, and the U.S.
Department of Energy Office of Science. The full acknowledgement can be found here: http://www.sdss3.org.

\end{document}